\documentclass[aps,prx,superscriptaddress,twocolumn]{revtex4-2}

\usepackage{amsmath}
\usepackage{graphicx}
\usepackage{hyperref}
\usepackage{braket}
\usepackage{color}
\usepackage{xcolor}
\usepackage{array}
\usepackage{colortbl}
\usepackage{makecell}
\usepackage{fancyhdr} 
\usepackage[textwidth=1.5cm,textsize=tiny]{todonotes}

\newcommand{\param}[1]{\emph{#1}}   
\newcommand{\val}[1]{\texttt{#1}}     

\begin{document}

\title{Optimized Matrix-Product State Simulations of Quantum Error Correction Circuits}

\author{Asier Pi\~neiro Orioli}
\affiliation{QPerfect SAS, European Center for Quantum Sciences, 23 rue du Loess, Strasbourg, 67200, France}
\author{Chen Zhao}
\affiliation{QuEra Computing Inc., Boston, Massachusetts 02135, USA}
\author{Guido Masella}
\affiliation{QPerfect SAS, European Center for Quantum Sciences, 23 rue du Loess, Strasbourg, 67200, France}
\author{Tommaso Macr\`i}
\affiliation{QuEra Computing Inc., Boston, Massachusetts 02135, USA}
\author{Hengyun Zhou}
\altaffiliation{Current affiliation: Department of Electrical Engineering and Computer Science, Massachusetts Institute of Technology, Cambridge, MA 02139, USA}
\affiliation{QuEra Computing Inc., Boston, Massachusetts 02135, USA}
\author{Shannon Whitlock}
\affiliation{University of Strasbourg and CNRS, CESQ-ISIS (UMR 7006), Strasbourg, France}
\affiliation{QPerfect SAS, European Center for Quantum Sciences, 23 rue du Loess, Strasbourg, 67200, France}

\date{\today}

\begin{abstract}
Simulating quantum error correction (QEC) circuits including non-Clifford gates at scale is important to accelerate progress toward fault-tolerant quantum computing.
Here we demonstrate that matrix product state (MPS) techniques can handle many QEC circuits \emph{exactly} and without restriction on gate types. Crucially, we find that MPS efficiency depends sensitively on implementation choices, and we introduce a series of targeted optimizations that reduce bond dimensions and simulation time by several orders of magnitude compared to naive approaches. We illustrate this with examples including: (a) a rotated surface code quantum memory up to distance 11, (b) logical Bell-state preparation up to distance 9, (c) a 15-to-1 magic-state distillation circuit including hundreds of QEC rounds that we optimize to be simulated with only 11 logical qubits (187 physical qubits) and a maximal bond dimension of 64 in under 40 seconds, and (d) a narrow, deep random circuit that scales linearly with the number of T gates.
These results demonstrate the importance of circuit-level optimizations and position MPS as a valuable complement to near-Clifford simulators for QEC circuits.
\end{abstract}

\maketitle

\section{Introduction}

Quantum error correction (QEC) is fundamental for suppressing errors in quantum computers to low enough levels to enable useful applications~\cite{Shor1995, Steane1996, gottesman2009introductionqec, RaussendorfHarrington2007, Dalzell2025, babbush2025grandchallenge}.
Recent advances in quantum hardware have enabled the first experimental implementations of QEC codes and small logical circuits~\cite{Krinner2022, Ryan2021, Acharya2025, Bluvstein2024, SalesRodriguez2025, reichardt2025ftqc, paetznick2024logqubits, computing2026toric}, while theoretical efforts have reduced the resources required for QEC by orders of magnitude~\cite{beverland2022assessing, gidney2025factor2048bitrsa, Bravyi2024bicycle, webster2026pinnacle, yang2026rasql, cain2026shor, Katabarwa2024, Zhou_2025}.
Testing new proposals against noise requires simulation, which is often accomplished using efficient Clifford simulators such as Stim~\cite{Gidney2021Stim}.
However, assessing the impact of realistic errors, such as coherent or correlated noise~\cite{Aharonov_2006, Bravyi2018coherenterrors, Huang_2019, Iverson_2020, Poulin_PRL2017, barone2025colorcode}, and non-Clifford operations, such as magic state distillation~\cite{bravyi2005universalquantum, Bravyi_2012, Litinski_2019} and cultivation~\cite{gidney2024magicstatecultivationgrowing, sahay2025foldtransversal}, requires methods capable of simulating universal circuits.

Simulation complexity of quantum circuits depends loosely speaking on (at least) two axes: entanglement and magic.
Since universal statevector simulators are typically limited to up to 30-40 qubits, a number of stabilizer-based non-Clifford simulators have been proposed based on extended stabilizer-rank methods~\cite{bravyi2016improvedclassical, Bravyi2019simulationofquantum, surti2026magicstateprep} recently combined with ZX-calculus compression~\cite{Kissinger_2022, Sutcliffe_2025, haenel2026tsim}, extended tableau methods~\cite{yoder2012generalization, liSOFTHighperformanceSimulator2025, chaseClifftFastExact2026, fang2026symftuniversal}, quasiprobability methods~\cite{pashayan2015quasiprobs}, circuit cutting~\cite{perlin2023circuitcutting},
Pauli propagation~\cite{rall2019paulipropagation,rudolph2026paulipropagation, queraPPVM2026}, or sparse Pauli-frame representations~\cite{tuloupComputingLogicalError2026}. These methods can handle high-entangled states, but are typically limited by the amount of magic produced in the circuit, as quantified for example by the number of T gates, the stabilizer rank, or by the stabilizer nullity~\cite{Beverland_2020}, depending on the method.

An alternative approach is provided by tensor network methods~\cite{Cirac_RMP2021}, which are the gold standard in large-scale quantum circuit simulation~\cite{tindall2024IBMtensornetwork, pan2022sycamoresimulation, tindall2026dwavesim, leonteva2025comparative}.
The key advantage of tensor networks is their ability to handle large universal circuits including high magic, without limitations on gate types, instead being limited by the amount of entanglement generated (the metric depends on the tensor network). Promising ideas to combine tensor networks with stabilizer-based methods have been recently proposed~\cite{Masot2024stabilizertensor, mello2024hybridstabsmpos, lami2024quantumstatedesignsclifford, qian2024camps}; however, the use of tensor networks to directly simulate QEC problems remains under-explored~\cite{Poulin_PRL2017, manabe2025leakageerrors, barone2025colorcode}.
Among tensor networks, matrix product states (MPS)\cite{vidal2004efficientsimulation,schollwock2011dmrgmps,Paeckel_AP2019,Cirac_RMP2021} are particularly attractive due to their simplicity, efficiency and versatility.

Commonly, MPS techniques are understood to: (1) only work for 1D circuits, (2) require low entanglement, and (3) always lead to approximations.
However, these claims are often misleading. Many relevant quantum circuits can be efficiently represented \emph{exactly} with an MPS of finite, moderate bond dimension despite not being intrinsically 1D~\cite{Zhou_2020limits, Niedermeier_2024, Stoudenmire_2024, leonteva2025comparative}, even for large entanglement, magic and non-Gaussianity~\cite{deger2026efficiently}; and in other cases, close enough to exact~\cite{tindall2024IBMtensornetwork}.
While ``low'' entanglement is a rather subjective term, MPS can be useful for finite-size applications which might popularly be thought of as moderately entangled (e.g.~area law).
In the following, we provide further evidence to clarify these issues.

In this work, we demonstrate that MPS can simulate intermediate-scale error-corrected quantum circuits of practical interest in an exact regime, and including those with a considerable number of non-Clifford gates. A central result is that MPS efficiency is not a fixed property of the QEC circuit, but depends sensitively on circuit and MPS implementation choices. A naive implementation of these circuits is intractable, even for highly optimized simulators, whereas the right choices make exact simulation feasible.
To illustrate this we benchmark the performance of MPS using the rotated surface code of distance $d$ and four different circuit types: a memory circuit (up to $d=11$), a logical Bell state preparation circuit (up to $d=9$), a 15-to-1 magic state distillation (MSD) circuit (up to $d=5$), and a deep random circuit ($d=3$).
The $d=3$ MSD circuit comprises 187 physical qubits, hundreds of rounds of QEC, and 15 logical T gates, and can be simulated with unit fidelity in under 40 seconds; for $d=5$ it goes up to 539 physical qubits in around 75 minutes.
The random circuit is used to explicitly show that MPS scales linearly with depth at fixed bond dimension; in particular, T gates and other single-qubit non-Clifford gates are cheap for MPS as they do not generate entanglement.
These results are enabled by circuit-level optimizations, particularly qubit and gate ordering and ancilla handling, that reduce the required bond dimensions and drastically improve MPS runtime by several orders of magnitude compared to naive implementations.

The encoded circuits studied here have strong stabilizer structure and relatively low magic---while the physical qubit number is large, magic only appears at the logical qubit level, which is low. Accordingly, we have found that recent near-Clifford simulators such as PPVM~\cite{queraPPVM2026} or CAMPS~\cite{qian2024camps} exploit this very effectively; such that even after the optimizations developed below, MPS does not match their absolute runtime on these instances. This is strongly problem-dependent (Sec.~\ref{sec:discussion}), and we expect the techniques introduced here will still find value in other contexts or in further improving the performance of hybrid Clifford + MPS approaches.

\begin{figure}[t!]
    \centering
    \includegraphics[width=\columnwidth]{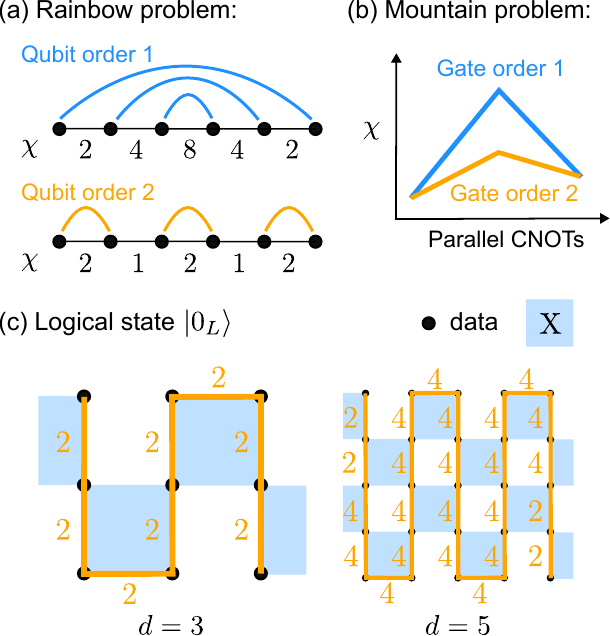}
    \caption{(a) The ``rainbow'' problem: entanglement shared across widely separated qubit pairs requires large bond dimensions under a naive ordering, but a simple reordering enforces nearest-neighbor connectivity and reduces $\chi_\text{max}$ to 2. (b) ``The mountain'' problem: two orderings of a commuting gate set produce identical final states but very different peaks in bond dimension during the circuit. (c) Optimal snake ordering for the state $\ket{0_L}$ of the rotated surface code for $d=3$ and 5. Light-blue squares represent the $X$ stabilizers of $\ket{0_L}$; the orange line represents the MPS qubit chain. The number next to each link gives the bond dimension $\chi_i$, determined by the number of stabilizers cut at that link, see Eq.~(\ref{eq:open_stabs}).}
    \label{fig:logstate}
\end{figure}

\section{Matrix-Product States for Quantum Error Correction\label{sec:mps_qec}}

Matrix Product States (MPS) are a type of tensor network that represent a high-dimensional tensor as a one-dimensional product of lower-rank tensors. An MPS representation of a quantum state is defined as~\cite{Cirac_RMP2021}
\begin{equation}
    \ket{\psi} = \sum_{s_1, s_2, \ldots} A^{s_1}_1 \cdot A^{s_2}_2 \cdot \ldots \cdot A^{s_N}_N \ket{s_1 s_2 \ldots} ,
\label{eq:mps}
\end{equation}
where $A^{s_k}_k$ are matrices of size $\chi_{k-1} \times \chi_k$ and the `$\cdot$' denotes matrix multiplication. 
Any quantum state can be represented as an MPS, but the efficiency of the representation depends on the size of the $A^{s_k}_k$ matrices, i.e.~on the \emph{bond dimensions} $\chi_k$. $\chi_k$ is the Schmidt rank entanglement between all qubits on the left, $i\leq k$, and on the right, $j>k$. Thus, MPS methods are most efficient for systems with low bipartite entanglement across any left-right system bipartition. 

Most MPS techniques, unlike Clifford simulators, do not explicitly exploit the structure and predominantly Clifford nature of many QEC circuits (see \cite{Masot2024stabilizertensor, mello2024hybridstabsmpos, lami2024quantumstatedesignsclifford, qian2024camps} for that line of research), so it is a priori unclear how well a general-purpose MPS simulator will perform on such problems. In this work, we identify two main implementation details which have a huge impact on performance, see Fig.~\ref{fig:logstate}:
\begin{enumerate}
    \item \textbf{Qubit ordering:} The entanglement structure of a state $\ket{\psi}$ is fixed; however, the bond dimensions $\chi_k$ depend on how qubits are mapped to the 1D tensor chain. Consider the ``rainbow'' problem in Fig.~\ref{fig:logstate}a: entanglement shared across widely separated qubit pairs (e.g.~Bell states) naively leads to large bond dimensions, but a simple reordering enforces nearest-neighbor connectivity and reduces the maximal bond dimension $\chi_\text{max}$ to 2. Since most QEC circuits lack a natural 1D layout, qubit ordering can have a large impact on performance.

    \item \textbf{Gate ordering:} A set of commuting gates, e.g.~$\{g_1, g_2\}$, has the same effect on the final state regardless of order, $g_1g_2\ket{\psi} = g_2g_1\ket{\psi}$; however, the transient states $g_1\ket{\psi}$ and $g_2\ket{\psi}$ can have very different entanglement properties. Since gates are applied sequentially in simulation, different orderings can produce very different peaks in bond dimension during the circuit (the ``mountain'' problem, Fig.~\ref{fig:logstate}b; see also Fig.~\ref{fig:cnot_bonddims} in Appendix) even when the final state is identical. For error correction circuits optimizing the order of parallel CNOTs can substantially reduce this peak, directly improving MPS efficiency.
\end{enumerate}

The main challenge of choosing the best implementation, e.g.~qubit and gate orderings, for a given circuit is that there are too many possibilities and few guidelines. We illustrate how $\chi_{max}$ depends on qubit ordering with a simple example for a single logical qubit. Following Ref.~\cite{Hamma_2005, zhao2026gentanglemententropy}, the bond dimension $\chi_k$ of a given bipartition for a CSS-type stabilizer state, i.e.~a state where all stabilizers are either products of $X$ or $Z$, is given by
\begin{equation}
    \chi_k = \mathrm{min} (2^{n_\text{open}}, 2^{\mathrm{min}(n_\text{left}, n_\text{right})}).
\label{eq:open_stabs}
\end{equation}
Here, $n_\text{open}$ is the \emph{minimal} number of ``open'' stabilizers of type $X$ (the formula also holds for $Z$ stabilizers), i.e.~stabilizers that have support on qubits on both sides of the bipartition, and $2^{\mathrm{min}(n_\text{left}, n_\text{right})}$ is an MPS bound set by the number of qubits on the left ($n_\text{left}$) and right ($n_\text{right}$) of the bipartition. By ``minimal number'' we mean the minimum over all possible equivalent sets of stabilizers.
While this can sometimes be easily computed by visual inspection, it can also be systematically computed as $n_\text{open} = r^Z_A+r^Z_B-r^Z_{AB}$, where $H_Z = (H_Z^A\ H_Z^B)$ is the parity check matrix for $Z$ stabilizers and $A$ and $B$ are the two sides of the bipartition \cite{zhao2026gentanglemententropy} (similar for $Z\leftrightarrow X$).

Based on the relationship of Eq.~(\ref{eq:open_stabs}), a heuristic to find good qubit and gate orderings is to minimize the number of open stabilizers at every bond, which minimizes the bond dimensions. Dynamically, this also means that whenever a stabilizer must be opened, it is better to break it over as few bonds or gates as possible. Consider the logical state $\ket{0_L}$ of a rotated surface code of distance $d$~\cite{Bombin_2007_rotated} with $d^2$ data qubits. To find a good qubit ordering, we focus on $d=3$ and the $X$ stabilizers (Fig.~\ref{fig:logstate}c). We build the MPS qubit ordering one step at a time, always choosing the next qubit to minimize the number of open stabilizers. The upper-left (or lower right) qubit is the best choice for qubit 1, as it touches only a single 2-qubit stabilizer, leading to $\chi_1=2$. The natural next choice is the only other qubit belonging to that same 2-qubit stabilizer. This closes it, while opening the adjacent 4-qubit stabilizer, keeping $\chi_2 = 2$. We then continue along that edge picking up the remaining qubits of the now-open 4-qubit stabilizer without opening any new ones, so $\chi_3 = \chi_4 = 2$. The center qubit closes this stabilizer while opening the next one, giving $\chi_5 = 2$.
At each step, the optimal move is to continue along the path that closes or partially closes an open stabilizer before venturing into a new one. Following this logic consistently leads naturally to the snake pattern of Fig.~\ref{fig:logstate}, with bond dimension $\chi_\text{max} = 2$ throughout.

The same logic extends to $d>3$ and to any of the four logical states $\ket{0_L}, \ket{1_L}, \ket{+_L}, \ket{-_L}$, yielding
\begin{equation}
\chi_\text{max} = 2^{(d-1)/2} \quad \text{(logical state)}.
\label{eq:chimax_logicalstate}
\end{equation}
Incidentally, this is consistent with the area-law entanglement of the surface code~\cite{Hamma_2005}.
Even a $d=11$ surface code logical qubit requires a bond dimension of only $\chi_\text{max}=32$ for exact representation. An arbitrary logical state $\alpha\ket{0_L}+\beta\ket{1_L}$ would require at most $2\chi_\text{max}$ with respect to Eq.~(\ref{eq:chimax_logicalstate}). Notice that for $\ket{\pm_L}$ the best ordering is the snake of Fig.~\ref{fig:logstate}c rotated by 90$^{\circ}$.

Until now, and to our knowledge, the best reported result was $\chi_\text{max}=4$~\cite{manabe2025leakageerrors} to represent the logical state $\ket{+_L}$ of the $3\times d$ rotated surface code.
The above method explains why: the snake ordering in Fig.~3 of \cite{manabe2025leakageerrors} (which includes ancillas) opens 2 stabilizers. For the purpose of just representing $\ket{+_L}$ we can improve upon this using the \val{north} ordering of Fig.~\ref{fig:memory}, which incorporates the ancillas and leads to $\chi_\text{max}=2$.

In the following sections, we apply this logic to several QEC circuits. However, circuits are intrinsically harder to optimize than static states: a circuit passes through many intermediate states for which the optimal qubit ordering may differ, and qubit ordering and gate ordering are interdependent and must be optimized jointly. Therefore, we explore these trade-offs through numerical simulations.

\section{Simulations\label{sec:simulations}}

To construct and run the circuits we use the MIMIQ\textsuperscript{TM} simulation framework by QPerfect~\cite{leonteva2025comparative}. In MIMIQ\textsuperscript{TM}, gates are applied to the MPS by representing them as Matrix Product Operators (MPOs)~\cite{Paeckel_AP2019}, where the analog of the bond dimension $\chi_k$ is called the entanglement dimension $D_k$. Unless stated otherwise, we apply gates as MPOs with $D_k \leq 2$, corresponding to the maximum operator Schmidt rank of a CNOT gate. Gates are thus applied one by one, except when multiple gates can be compressed into a single MPO without exceeding $D_k = 2$. This compression does not affect intermediate bond dimensions but can reduce simulation time. For MPO application we use the zip-up method~\cite{Paeckel_AP2019}, which prioritizes minimal bond dimensions over runtime, and ensures minimal $\chi_\text{max}$ representations at each step by applying an SVD compression sweep. Note that using $D_k>2$ can sometimes yield faster runtimes, but we exclude this for simplicity.

All simulation results are exact to high precision: specifically, $1-F \leq 10^{-8}$, where $F = |\braket{\psi|\psi_\text{MPS}}|^2$ is estimated from the discarded Schmidt weights. All simulations are performed on a single cluster node (Intel Xeon, Sapphire Rapids; 4 threads; 125 GiB RAM) and we note that higher thread parallelization can yield further speedups, especially for large bond dimensions. The runtimes given correspond to the minimum or mean runtime over several runs unless stated otherwise, and do not account for circuit construction nor code (pre-)compilation times. All circuits have been verified through logical tomography of the final state and detector behavior where appropriate.

We consider circuits with a Pauli noise model, applied using the quantum trajectories approach~\cite{Daley_2014}. Since Pauli operators do not change the Schmidt-rank entanglement, the effect of Pauli noise on simulation runtime is negligible (checked empirically). Of course, they could change entanglement after propagating through \textsc{T} gates; however, in our case the MSD circuit only has \textsc{T} gates at the end, and the random circuits already saturate bond dimensions at $2^{\lfloor N_L/2 \rfloor}$ at the logical level. We therefore benchmark simulation runtimes on noiseless circuits.

\section{Memory circuit\label{sec:memory}}

\begin{table*}[ht]
\centering
\begin{tabular}{|c|c|c|c|c|c|c|}
    \hline
    $d$ & $n_\text{qubits}$ & \param{qubit-order} & \param{qec-layer-order} & \param{qec-cnot-order} & $\chi_\text{max}$ & runtime (s) \\
    \hline
    3  & 17  & \val{northeast} & \val{N} & \val{southeast} & 4    & 0.0031 \\
    5  & 49  & \val{northeast} & \val{N} & \val{southeast} & 16   & 0.13   \\
    7  & 97  & \val{northeast} & \val{N} & \val{southeast} & 64   & 10.0   \\
    9  & 161 & \val{northeast} & \val{N} & \val{southeast} & 256  & 587    \\ 
    11 & 241 & \val{northeast} & \val{N} & \val{southeast} & 1024 & 42\,239  \\ 
    \hline
\end{tabular}
\caption{Results for the memory circuit. The logical qubit is prepared in $\ket{0_L}$, followed by $d-1$ rounds of QEC. We show the parameter combinations that were found to minimize $\chi_\text{max}$ and the runtime. Runtimes are the minimum runtimes over 10 repetitions for $d\leq7$ and a single run for $d=9$ and $d=11$. Simulations were done on a single cluster node (Intel Xeon Sapphire Rapids; 4 threads; 125 GiB), see also Sec.~\ref{sec:simulations}.}
\label{tab:memory_bench}
\end{table*}

\begin{figure}[t!]
    \centering
    \includegraphics[width=\columnwidth]{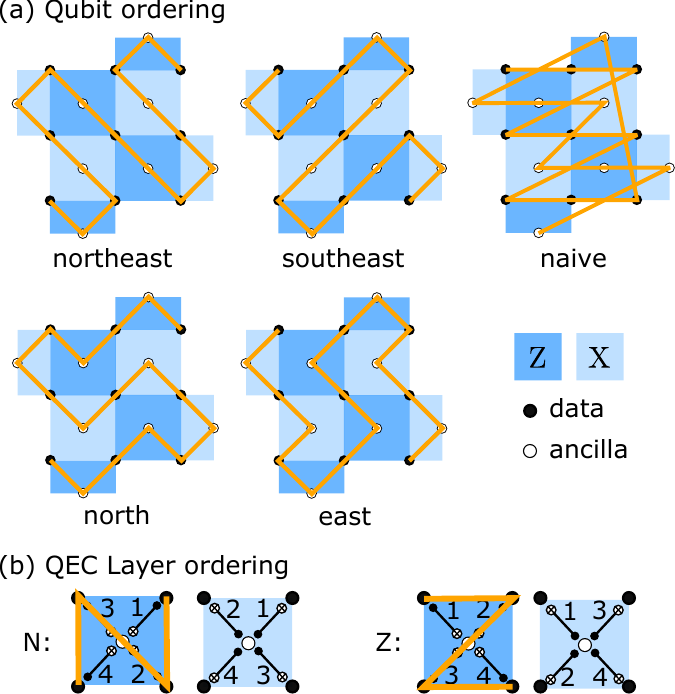}
    \caption{Parameter choices for memory circuit. (a) \param{qubit-order} options. The orange line in (a) represents the 1D MPS, where each qubit is a tensor. (b) \param{qec-layer-order} options. Each of the 4 QEC layers, $k\in\{1,2,3,4\}$, consists of all CNOTs between each stabilizer ancilla and the data qubit in the direction indicated by $k$. The orange line in (b) is just a visual aid.}
    \label{fig:memory}
\end{figure}

We first consider a memory circuit using a distance $d$ rotated surface code comprising $d^2$ data qubits and $d^2-1$ ancillas, as shown in Fig.~\ref{fig:memory}. The qubits are prepared in the logical state $\ket{0_L}$ by initializing all data qubits in state $\ket{0}$ and then measuring the $X$ stabilizers. This is then followed by $d-1$ full QEC rounds and a final measurement of the data qubits.
Each QEC round consists of initializing ancillas, 4 layers of parallel CNOTs and ancilla measurements.

The memory circuit presents a nontrivial MPS simulation task, distinct from the logical state preparation analyzed above. Although both involve preparing $\ket{0_L}$, the intermediate states traversed during a QEC round are not logical codewords: as each gate is applied, the stabilizer structure of the state evolves continuously, and each intermediate state may have a different optimal qubit ordering. Only at the end of a full QEC round does the state return to $\ket{0_L}$. The efficiency of the MPS representation therefore depends on both qubit ordering and gate ordering, making the optimization problem significantly more complex than in the static case. We address it using a combination of heuristics and brute-force computation.

To search for optimal configurations for the memory circuit, we consider three relevant parameters of the circuit (depicted in Fig.~\ref{fig:memory}): 
\begin{itemize}
	\item \textbf{\param{qubit-order} :}
	The physical qubit ordering is defined as a map $\sigma:[1,N]\rightarrow[1,N]$ that assigns to each of the $N$ physical qubits an index position in the MPS of Eq.~(\ref{eq:mps}). We consider 5 distinct orderings as shown in Fig.~\ref{fig:memory}a. The \val{north}/\val{east} snakes extend the optimal logical-state orderings found above to include ancilla qubits; the \val{northeast}/\val{southeast} variants follow diagonal paths. The \val{naive} ordering indexes data qubits 1 to $d^2$ followed by ancillas top to bottom, representing a natural output of circuit construction libraries, included for comparison.
    
	\item \textbf{\param{qec-layer-order} :}
    We implement a QEC round using 4 layers of parallel CNOTs, which can be ordered in different ways. Of those, 8 layer orderings are compatible with resilience to hook errors~\cite{Tomita_2014surfacecode}, the most significant distinction is whether or not the CNOTs of the second and third layers run parallel or orthogonal to the snake. We focus on two representative orderings, labeled \val{N} and \val{Z} (Fig.~\ref{fig:memory}b), as other orderings show similar performance for symmetry reasons.
    
	\item \textbf{\param{qec-cnot-order} :}
    Within each QEC layer, the $n$ parallel physical CNOTs can be ordered in $n!$ different ways. To reduce the space, we choose to order the physical CNOTs acting on pairs $(q^1_k, q^2_k)$ by increasing $\mathrm{min}(\sigma(q^1_k), \sigma(q^2_k))$, using each of the five \param{qubit-order} options above. While many more orderings are possible, we find a few are sufficient to illustrate the effect of this parameter on performance.
\end{itemize}

Table~\ref{tab:memory_bench} shows the results of an exhaustive search over all parameter combinations for $d=3$, which were then extended to up to $d=11$ using the best candidates. We empirically find that the memory circuit for $\ket{0_L}$ can be simulated most efficiently using \param{qubit-order}=\val{northeast} and \param{qec-layer-order}=\val{N} with an MPS of bond dimension
\begin{equation}
	\chi_\text{max} = 2^{d-1} \quad \text{(memory)}.
\label{eq:chimax_memory}
\end{equation}
Notably, the \val{east} ordering that is optimal for the state $\ket{0_L}$ itself is suboptimal for the memory circuit.
The minimum achievable bond dimension follows from the stabilizer structure mid-QEC-round (after the first two CNOT layers), which corresponds to that of the unrotated surface code~\cite{McEwen_2023}. This code has roughly twice as many stabilizers along each dimension, requiring a bond dimension equal to the square of the logical-state result of Eq.~\ref{eq:chimax_logicalstate}.

For \param{qubit-order}=\val{northeast} and \param{qec-layer-order}=\val{N}, $\chi_\text{max}$ is insensitive to \param{qec-cnot-order}, though it affects runtime through small differences in intermediate entanglement. The qubit orderings other than \val{northeast} lead to larger bond dimensions and runtimes, with the gap growing with distance $d$. The \val{naive} ordering is particularly costly: at $d=3$ it already requires $\chi_\text{max} = 64$ and becomes intractable at $d=5$. Perhaps more surprisingly, alternative geometrically natural snake-like orderings also perform significantly worse than \val{northeast}.
For example, \val{east} and \val{southeast} both require $\chi_\text{max}=8$ at $d=3$ and $\chi_\text{max}=32$ at $d=5$ (i.e.~scaling as $2^{d}$), whereas \val{north} requires $\chi_\text{max}=16$ and 64 (i.e.~scaling as $2^{d+1}$), respectively.
The growth in bond dimension required translates into longer runtimes as the SVD compression scales as $\chi^3$.
This shows that small implementation choices can lead to orders-of-magnitude differences in performance: the best $d=3$ choice is 100 times faster than \val{naive} and around 4 times faster than the worst snake-like ordering tried (around 25 times at $d=5$), see Fig.~\ref{fig:memory_runtimes} in the Apppendix for a wider overview.

We note that the optimal parameters for the memory circuit starting at $\ket{+}$ are the same as in Table~\ref{tab:memory_bench} but rotated by 90 degrees, i.e.~\param{qubit-order}=\val{southeast}, \param{qec-layer-order}=\val{Z}, and \param{qec-cnot-order}=\val{northeast}.

\section{Bell state circuit\label{sec:bell}}

\begin{table*}[ht!]
\centering
\begin{tabular}{|c|c|c|c|c|c|c|c|}
    \hline
    $d$ & $n_\text{qubits}$ & \param{cnotL-cnot-order} & \param{qubit-order} & \param{qec-layer-order} & \param{qec-cnot-order} & $\chi_\text{max}$ & runtime (s) \\
    \hline
    3 & 34  & \val{north}      & \val{north}, \val{east} & \val{Z}, \val{Z} & \val{north}, \val{southeast} & 8   & 0.018  \\
    5 & 98  & \val{north}     & \val{north}, \val{east} & \val{Z}, \val{N} & \val{east}, \val{north}          & 32  & 1.38   \\
    7 & 194 & \val{north}      & \val{north}, \val{east} & \val{Z}, \val{N} & \val{east}, \val{north}           & 128 & 152  \\
    9 & 322 & \val{north}      & \val{north}, \val{east} & \val{Z}, \val{N} & \val{east}, \val{north}          & 512 & 14202   \\ 
    \hline
\end{tabular}
\caption{Results for the Bell state circuit. The logical qubits are prepared in $\ket{+_L 0_L}$, followed by 1 round of QEC and 2 transversal logical CNOT gates. We show results for the optimal (north, east) qubit ordering. Runtimes are the minimum runtimes over 10 repetitions for $d\leq5$ and a single run for $d\geq7$.  Simulations were done on a single cluster node (Intel Xeon Sapphire Rapids; 4 threads; 125 GiB), see also Sec.~\ref{sec:simulations}.}
\label{tab:bell_bench}
\end{table*}

\begin{figure}[t!]
    \centering
    \includegraphics[width=\columnwidth]{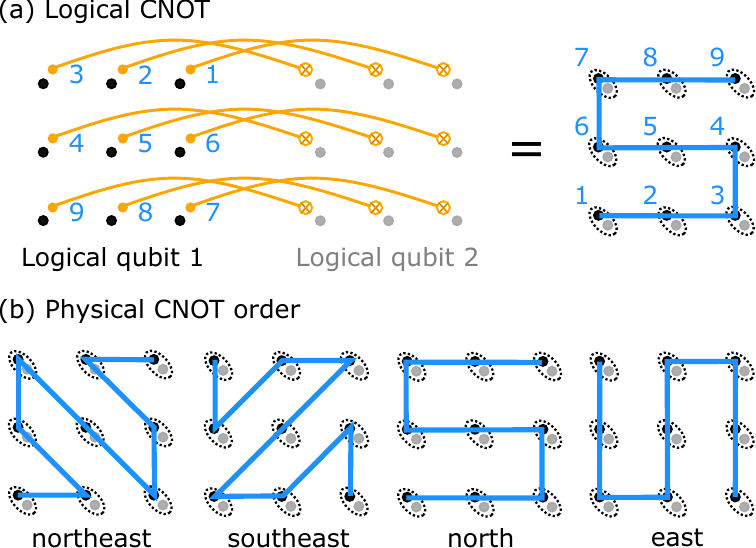}
    \caption{(a) Transversal logical CNOT between two logical qubits. The order of physical CNOTs given by the blue numbers is represented by a snake on the right (corresponding to \val{north}). (b) Parameter choices for the \param{cnotL-cnot-order} parameter. The indices grow in the compass direction indicated by the name.}
    \label{fig:bell}
\end{figure}

As a second case study, we consider a circuit that prepares a logical 
Bell state using two logical qubits. The circuit has two parts: (i) a memory 
subcircuit preparing $\ket{+_L 0_L}$ with 2 QEC rounds (the first round measures 
only the non-deterministic stabilizers, as before), followed by (ii) two transversal 
logical CNOTs between all data qubits of the two logical qubits, see 
Fig.~\ref{fig:bell}. This creates a logical Bell state and then brings the logical qubits back to their original product state. We include a second QEC round in the memory subcircuit because the intermediate bond dimensions during the first round are smaller and not 
representative of the full circuit.

The optimal MPS parameters for this circuit reflect a competition between two 
regimes. During the memory subcircuit, the \val{northeast} (\val{southeast}) qubit ordering 
minimizes bond dimensions for $\ket{0_L}$ ($\ket{+_L}$), as established in 
Sec.~\ref{sec:mps_qec}. The transversal CNOT part, however, acts directly on the 
logical state $\ket{+_L 0_L}$, for which the best ordering is (\val{north}, \val{east}), where the notation (\val{val}$_1$, \val{val}$_2$) denotes that logical qubit 1 has parameter value \val{val}$_1$ and logical qubit 2 has \val{val}$_2$. We resolve this competition through a combination of numerical 
search and analytical arguments.

Since the circuit involves two logical qubits, each can be assigned an independent \param{qubit-order}, \param{qec-layer-order} and \param{qec-cnot-order}. Since entanglement between logical qubits is generated only by the logical CNOT, we assume the qubits of each logical qubit are ordered in contiguous blocks, i.e.~$\sigma(i) < \sigma(j)$ where $i\in S_1$ and $j\in S_2$ correspond to physical qubit index labels of logical qubits 1 and 2, respectively, and $\sigma$ is the qubit ordering map. In addition, we consider one more parameter specific to this circuit:
\begin{itemize}
    \item \textbf{\param{cnotL-cnot-order}:} The $d^2$ 
    parallel physical CNOTs implementing each logical CNOT can be ordered in $d^2!$ ways. Each logical CNOT can at most double 
    the bond dimension between the two logical qubit blocks (assuming they start in 
    a logical state). However, as in the memory circuit, intermediate states can 
    transiently require higher bond dimensions depending on the order in which the 
    physical CNOTs are applied. We consider 4 orderings: \val{north}, \val{east}, \val{southeast}, 
    and \val{northeast}, shown in Fig.~\ref{fig:bell}, justified below.
\end{itemize}

To find the best parameters we again perform an exhaustive search for $d=3$ and 
extend the best candidates to higher $d$ (Tab.~\ref{tab:bell_bench}). We empirically find that 
the Bell state circuit can be simulated exactly with
\begin{equation}
    \chi_\text{max} = 2^d \quad \text{(Bell)}.
\label{eq:chimax_bell}
\end{equation}
To understand this it is instructive to analyze the most important optimization knobs:

\paragraph{cnotL-cnot-order.}
The minimal $\chi_\text{max}$ of Eq.~(\ref{eq:chimax_bell}) is achieved for both the \val{north} and \val{east} CNOT orderings (Fig.~\ref{fig:bell}).
To understand why, we first introduce a proxy for overall bond dimension growth: $\chi_\text{inter}$, the bond dimension across the link between the two logical qubit blocks, i.e.~between the last qubit of logical qubit 1, $\mathrm{max}_{i\in S_1}\sigma(i)$, and the first qubit of logical qubit 2, $\mathrm{min}_{j\in S_2}\sigma(j)$.

Each physical CNOT propagates $X$ operators from the first logical qubit to the second, extending stabilizers of the first block across the inter-block cut and doubling $\chi_\text{inter}$ for each open stabilizer. The optimal \param{cnotL-cnot-order} therefore minimizes the number of simultaneously open stabilizers, analogously to the qubit ordering principle of Sec.~\ref{sec:mps_qec}. This gives 
a peak growth of
\begin{equation}
    \chi_\text{inter} \rightarrow 2^{(d+1)/2}\,\chi_\text{inter}
\label{eq:chi_inter}
\end{equation}
during logical CNOT application.
To see why, note that the first logical qubit starts in $\ket{+_L}$, stabilized by the $(d^2-1)/2$ code $X$-stabilizers and the logical $X_L$. A stabilizer $s_1$ 
contributes to $\chi_\text{inter}$ only while CNOTs are being applied to its support qubits; once the last such CNOT is applied, $s_1$ transforms into $s_1 s_2$ 
(where $s_2$ is the corresponding stabilizer on the second logical qubit) and can be closed by multiplication with $s_2$. \val{north} and \val{east} orderings minimize the number of simultaneously open stabilizers and achieve $2^{(d+1)/2}$ (at worst), while \val{southeast} and \val{northeast} orderings open more stabilizers concurrently and lead to a larger transient $\chi_\text{inter}$, see Fig.~\ref{fig:cnot_bonddims} in the Appendix for a finer-grained comparison.

\paragraph{qubit-order.}
The (\val{north}, \val{east}) heterogeneous qubit ordering achieves Eq.~(\ref{eq:chimax_bell}) across all distances in Tab.~\ref{tab:bell_bench}, consistent with the optimal orderings 
for the respective logical states found in Sec.~\ref{sec:mps_qec}:
\val{north} for $\ket{+_L}$ and \val{east} for $\ket{0_L}$.
To see why this also minimizes the bond dimension during the logical CNOT, notice that each physical CNOT multiplies all bond dimensions between its control and target by at most a factor 2.
Thus, the peak bond dimension during the logical CNOT is at most $\chi_\text{max}(\ket{+_L 0_L}) \times 2^{(d+1)/2}$, where the factor $2^{(d+1)/2}$ is the peak growth of $\chi_\text{inter}$ in Eq.~(\ref{eq:chi_inter}). 
The (\val{north}, \val{east}) ordering minimizes $\chi_\text{max}(\ket{+_L 0_L}) = 2^{(d-1)/2}$ 
[Eq.~(\ref{eq:chimax_logicalstate})], directly yielding Eq.~(\ref{eq:chimax_bell}) above.

Although (\val{north}, \val{east}) is suboptimal for the memory subcircuit, it incurs only a factor-of-2 overhead relative to the memory optimum of Eq.~(\ref{eq:chimax_memory}), 
reaching $\chi_\text{max} = 2^d$ there as well (\val{north} for $\ket{+_L}$ and \val{east} for $\ket{0_L}$)---equal to the $\chi_\text{max}$ reached during the logical CNOT. In contrast, the (\val{southeast}, \val{northeast}) qubit ordering, which is optimal for the memory subcircuit, achieves $\chi_\text{max}(\ket{+_L 0_L}) = 2^{d-1}$ (c.f.~Fig.~\ref{fig:logstate} for stabilizers opened) and is thus penalized heavily by the logical CNOT: multiplying by the factor $2^{(d+1)/2}$ yields $\chi_\text{max} = 2^{(3d-1)/2}$, far exceeding Eq.~(\ref{eq:chimax_bell}).
The (\val{north}, \val{east}) ordering, on the other hand, resolves the competition between the two regimes at no net cost.

We note that all $\chi_\text{max}$ reported increase by a factor of 2 if a QEC round is performed in between the two logical CNOTs, since the Bell state has twice the $\chi_\text{max}$ compared to $\ket{+_L0_L}$.

\paragraph{Remaining parameters.}
The \param{qec-layer-order} and \param{qec-cnot-order} parameters affect only the memory subcircuit, where the bond dimension is $2^d$ rather than the optimal $2^{d-1}$ of Eq.~(\ref{eq:chimax_memory}).
While these parameters did not affect the optimal bond dimension of Eq.~(\ref{eq:chimax_memory}), the suboptimal $2^d$ bound for \param{qubit-order}=\val{east} and \val{north} is only achieved for some values of \param{qec-layer-order} and \param{qec-cnot-order}; an example is shown in Tab.~\ref{tab:bell_bench}.
The optimal choices enable exact simulation of the Bell circuit up to $d = 9$ with $\chi_\text{max} = 512$.

\section{Magic State Distillation circuit\label{sec:msd}}

\begin{figure}
    \centering
    \includegraphics[width=\columnwidth]{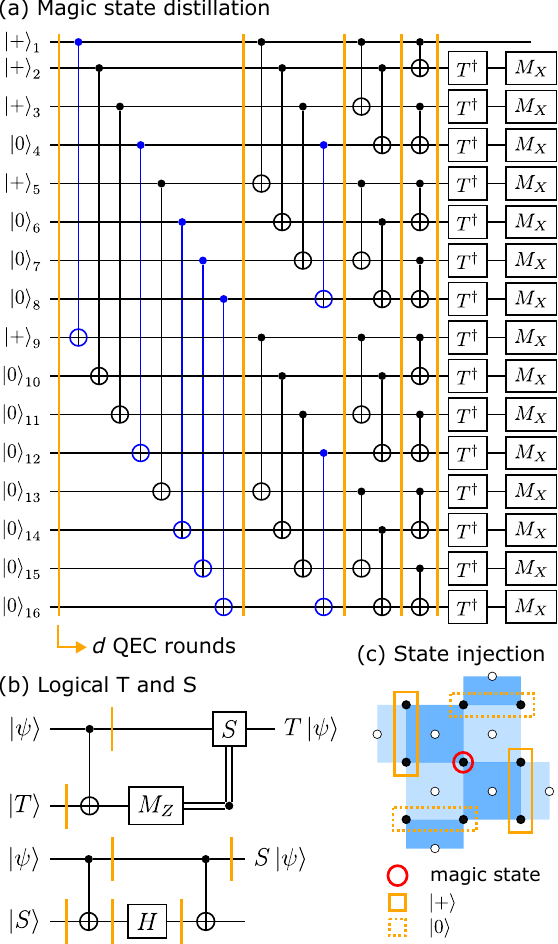}
    \caption{(a) 15-to-1 magic state distillation circuit from Ref.~\cite{beverland_cost_2021}. The blue CNOTs act trivially on the state and can be removed. (b) Implementation of $T$ gates through teleportation with an ancilla prepared in the magic state $\ket{A_L}=(\ket{0_L}+e^{i\pi/4}\ket{1_L})/\sqrt{2}$, and S gate teleportation-type gadget from Ref.~\cite{Fowler_PRA2012} using an ancilla prepared in $\ket{Y_L}=(\ket{0_L}+i\ket{1_L})/\sqrt{2}$. (c) Magic state injection protocol from Ref.~\cite{LaoCriger_Proc2022}: data qubits are prepared in the states indicated and two QEC rounds are performed right after. The protocol can be used for both $S$ and $T$ injections.}
    \label{fig:msd_circuit}
\end{figure}

The insights on MPS optimization gained for the memory and Bell circuits can be used to aid the optimization of larger circuits. 
As an example, we consider a logical circuit implementing a 15-to-1 magic state distillation (MSD) circuit \cite{beverland_cost_2021,bravyi2005universalquantum,Haah2018codesprotocols} using the rotated surface code as shown in Fig.~\ref{fig:msd_circuit}.
For variety, we also consider the 3-to-1 and 7-to-1 subcircuits which are given by the circuit of Fig.~\ref{fig:msd_circuit} constrained to the first 4 and 8 qubits and with $T$ gates substituted by $Z$ and $S$, respectively (this results in $Z$ and $S$ state distillation~\cite{wan2026msdnumerics}).

As before, logical qubits are initialized by measuring the stabilizers, and logical CNOTs are implemented transversally. The CNOTs that act trivially on the state are skipped (marked in blue for the 15-to-1 case, Fig.~\ref{fig:msd_circuit}a). We perform $d$ rounds of QEC after every operation, as specified in Fig.~\ref{fig:msd_circuit}a-b by orange lines.
T gates use state injection and gate teleportation~\cite{Fowler_PRA2012}. S gates are implemented as in Fig.~\ref{fig:msd_circuit}, omitting the final H gate~\cite{Fowler_PRA2012} and discarding the ancilla \footnote{While more efficient fold-transversal methods exist for the S gate~\cite{chen2024transversalclifford, Moussa_2016}, we choose this method to simplify the optimization parameter space.}. State injection for logical S and T states follows the mid-register protocol of Ref.~\cite{LaoCriger_Proc2022} (Fig.~\ref{fig:msd_circuit}). Hadamard gates involve transversal H operations followed by a code rotation implemented via SWAP gates.
The final optimized circuit for $d=3$ contains 187 physical qubits (see below), around 30000 physical operations and hundreds of QEC rounds.

The MPS optimization of this circuit can be divided into two parts: logical and physical levels. Both can be optimized independently.

\subsection{Logical level}

At the logical level there are several knobs that can be tuned: the logical qubit order, the order of operations (\textsc{CNOT}, \textsc{T}, \textsc{S}, \textsc{M}$_\textsc{X}$), and the number and order of logical ancillas for \textsc{T} and \textsc{S} gates.
The parameters we found to have the biggest impact are the following:
\begin{itemize}
    \item \textbf{\param{ancilla-placement}:} Each \textsc{T} and \textsc{S} gate requires a logical ancilla to be executed, but these ancilla can be reused. Thus, the number of total ancillas and their position in the MPS can be optimized. We consider from 1 to $N_L-1$ ancillas ($N_L\in\{4,8,16\}$ is the number of data qubits), positioned either at the end or symmetrically interspersed around the center.
    Using more ancillas has the advantage that each ancilla can be placed in the MPS close to the logical qubit it will interact with, at the cost of extra qubits. In contrast, having fewer ancilla reduces the qubit number at the cost of longer-range gates in the MPS.
    
    \item \textbf{\param{measure-early}:} Measurements can be pushed to earlier times since they commute with several operations. This can help reduce entanglement by projecting some qubits earlier. Specifically, one can measure the logical qubits that do not require an \textsc{S} gate correction before the other \textsc{S} gates are applied. The advantage is that subsequent \textsc{S} gates can then be applied on a less entangled state.
    
    \item \textbf{\param{depth-reorder}:} The MSD circuit can be written as a Directed Acyclic Graph (DAG) that can be traversed in different ways. Switching from breadth first to depth first corresponds to a space-time tradeoff: some qubits can be measured before others are even initialized. For the depth-first approach, one must also choose a priority order for the qubits. We consider: \val{none} (no reordering), \val{ascending} (1,\ldots,$N_L$), \val{descending} ($N_L$,\ldots,1), \val{alternating} ($N_L$, 1, $N_L-1$, 2,\ldots), or \val{center} ($N_L/2$, $N_L/2-1$, $N_L/2+1$,\ldots).

    \item \textbf{\param{qubit-reuse}:} When a qubit is measured, its slot can be used for either ancilla or other data qubits that have not been initialized yet (as in the depth-first approach of the previous point). This is useful to reduce total qubit numbers. We consider \val{false} (no reuse) and \val{true} (qubits are allocated as they come alive, more details below).
\end{itemize}

Of all the above choices we found the depth-first approach with \param{depth-reorder}=\val{descending} or \val{ascending} to be the most important optimization, as it allows to reduce the bond dimension needed to represent the state of the logical circuit by a factor 2 with respect to \val{none} for $N_L=8, 16$, specifically:
\begin{equation}
    \chi_\text{max} = N_L / 2\qquad \text{(Logical MSD)}.
\label{eq:chimax_msd_log}
\end{equation}
The order of operations in this case is such that the state prepared by the full MSD circuit before measurement (Bell state of the Reed-Muller code with the first qubit) is never fully constructed.
Combined with \param{qubit-reuse}=\val{true} and a single ancilla, the MSD circuits can be simulated with 3, 7 and 11 logical qubits for the $N_L=4, 8$ and 16 cases, respectively (see Figs.~\ref{fig:msd4_compressed}, \ref{fig:msd8_compressed}, and \ref{fig:msd16_compressed} in the Appendix), thus significantly reducing the qubit count and improving runtimes. For example, we observe around 40 times faster runtime for the $N_L=16$ $d=3$ case simulated in the next section compared to \param{depth-reorder}=\val{none} and \param{qubit-reuse}=\val{false}. Incidentally, such modifications can also be used in real quantum computer implementations where space is more scarce than circuit depth.

Regarding the qubit and ancilla ordering for the \param{depth-reorder}=\val{descending} method with \param{qubit-reuse}=\val{true}, we follow a simple procedure of dynamically adding qubits only as they are needed. When a qubit is measured, a slot is freed that will be used by the next fresh qubit needed. Specifically, data qubits are allocated at the smallest available index, and ancillas as close as possible to the data qubit they will interact with. While some further optimizations might be possible to slightly reduce the CNOT connectivity, we find this to be efficient enough for our purposes.

It is worth noting that with \param{depth-reorder}=\val{none} and \param{qubit-reuse}=\val{false} \footnote{This may be needed if a decoder requires decoding all $T$ teleportation ancilla measurements before the final measurement.}, the \param{ancilla-placement} and the \param{measure-early}=\val{true} parameters lead to significant improvements.
In this case, we find a single ancilla placed in the center to be the simplest and best solution, as it reduces the span of the CNOTs involved in it.

Other optimizations we attempted proved irrelevant.
Optimizing logical data qubit order can potentially reduce long-range gates, but we found no significant improvement with respect to the natural top-to-bottom ordering given in Fig.~\ref{fig:msd_circuit}. This is probably because the most expensive CNOTs in the MSD circuit are the last ones, which are already nearest-neighbor in the natural order.
Similarly, the logical CNOTs inside each layer of parallel logical CNOTs in the MSD circuit can be ordered in different ways as in previous sections, but we found no significant improvement compared to a standard top-to-bottom ordering, probably because of the hypercube CNOT symmetry of the circuit.

\subsection{Physical level}

\begin{table*}[ht!]
\centering
\begin{tabular}{|c|c|c|c|c|c|c|c|c|c|c|}
    \hline
    $N$ & $d$ & $n_\text{qubits}$ & \param{depth-reorder} & \param{qubit-reuse} & \param{cnotL-cnot-order} & \param{qubit-order} & \param{qec-layer-order} & \param{qec-cnot-order} & $\chi_\text{max}$ & runtime (s) \\
    \hline
    4 & 3 & 51 & \val{descending} & \val{true} & \val{north} & \val{southeast} & \val{Z} & \val{northeast} & 16 & 0.117 \\
    8 & 3 & 119 & \val{descending} & \val{true} & \val{north} & \val{southeast} & \val{Z} & \val{northeast} & 32 & 3.23 \\
    16 & 3 & 187 & \val{descending} & \val{true} & \val{north} & \val{southeast} & \val{Z} & \val{northeast} & 64 & 38.49 \\ \hline
    4 & 5 & 147 & \val{descending} & \val{true} & \val{north} & \val{southeast} & \val{Z} & \val{northeast} & 128 & 17.9 \\
    8 & 5 & 343 & \val{descending} & \val{true} & \val{north} & \val{southeast} & \val{Z} & \val{northeast} & 256 & 607 \\
    16 & 5 & 539 & \val{descending} & \val{true} & \val{north} & \val{southeast} & \val{Z} & \val{northeast} & 512 & 4506 \\
    \hline
\end{tabular}
\caption{Runtimes for the MSD circuit for $d=3$ and 5 rotated surface code. We show results for the optimal parameters. Runtimes are the minimum runtimes for $N_L\leq 8$ and the mean for $N_L=16$ because of the teleportation gadget (the $d=5$ $N_L=8, 16$ results are a single run though). Simulations done on a single cluster node (Intel Xeon Sapphire Rapids; 4 threads; 125 GiB), see also Sec.~\ref{sec:simulations}.}
\label{tab:msd_bench}
\end{table*}

Optimizing MPS for the physical-level circuit requires re-optimizing all the parameters we introduced for the memory and Bell circuits. Since the MSD circuit mostly consists of QEC rounds and logical CNOTs and these two components favor different orderings, the optimal parameters are not obvious. 
One would expect the optimal parameters found for the Bell circuit to be favored, since that circuit includes CNOTs and QEC rounds---but things can change in a more complex circuit, as we will see.

We order physical qubits in a hierarchical way: first we order the logical qubits according to the previous section, and then the physical qubits inside each of them, such that if $i_m$ and $j_n$ belong to logical qubits $m, n$ with $m>n$ then $\sigma(i_m)>\sigma(j_n)$.
Even though every single logical qubit can in principle use different \param{qubit-order} values, for simplicity we fix the same orderings for all of them.
The same applies to the \param{qec-cnot-order} and \param{qec-layer-order} parameters.

Since the MSD circuit is rather large and there are many parameters to optimize, we run a comprehensive study on the smaller 3-to-1 and 7-to-1 instances of the circuit, and then choose the best parameter combinations to run the full circuit.
We also use the teachings of the previous sections to select the most promising parameter combinations for the smaller circuits. Thus, we consider (\val{northeast}, \val{southeast}, \val{east}, \val{north}) for \param{qubit-order} and \param{qec-cnot-order}, (\val{N}, \val{Z}) for \param{qec-layer-order}, and (\val{east}, \val{north}) for \param{cnotL-cnot-order}.

Table~\ref{tab:msd_bench} summarizes the best runtimes found along with the optimal parameters used. Remarkably, the full MSD circuit with $d=3$ can be simulated in under 40 seconds and using less than 3Gb of memory. Both logical and physical-level optimizations are essential to obtain these results.

The minimal $\chi_\text{max}$ to simulate the physical MSD circuit depends on $N_L$ and $d$ in a subtle way. For $N_L=4$ we empirically find
\begin{equation}
    \chi_\text{max} = 2^{d+1} \quad
        (\text{Physical MSD}, N_L=4),
\label{eq:chimax_msd_phys_nl4}
\end{equation}
whereas for $N_L=8$ or 16 we find
\begin{equation}
    \chi_\text{max} = 
        N_L 2^{\text{min}[(3d-5)/2,\,d]} \quad (\text{Physical MSD}, N_L=8,16).
\label{eq:chimax_msd_phys_nl816}
\end{equation}
We checked these results up to $d=5$.
For lower $d$ we find \param{qubit-order}=\val{southeast} leads to a physicalization factor $2^{(3d-3)/2}$ on top of the logical $N_L/2$ of Eq.~(\ref{eq:chimax_msd_log})---it is limited by logical CNOTs. The factor corresponds to half of a logical CNOT's $\chi_\text{max}$ with the same parameters ($2^{d-1}$ to represent the logical state and $2^{(d+1)/2}$ for the logical CNOT, Eq.~(\ref{eq:chi_inter})). The factor of 2 improvement we attribute to an accident of the particular state created~\footnote{In general, for $d=3$ a logical CNOT leads to a factor of 4 increase in transient bond dimensions according to Eq.~(\ref{eq:chimax_msd_log}), but for the MSD circuit we observe an occasional factor of 2 instead, typically when the CNOT acts on qubits that are already entangled with other qubits.}, c.f.~Sec.~\ref{sec:random}.
For higher $d$ we find \param{qubit-order}=\val{north} leads to a physicalization factor $2^{d+1}$ on top of $N_L/2$---it is limited by QEC rounds. The factor corresponds to the $\chi_\text{max}$ required for QEC rounds for arbitrary logical states ($\ket{0_L}$ or $\ket{+_L}$) for this qubit order~\footnote{While the logical CNOTs would also require $2^{d+1}$ ($2^{(d+1)/2}$ for an arbitrary logical state and $2^{(d+1)/2}$ for the logical CNOT $2^{(d+1)/2}$, Eq.~(\ref{eq:chi_inter})), we again observe a factor of 2 smaller values probably because of the particular state created.}.
The case $N_L=4$ is an exeption where we find the QEC rounds require half this factor, yielding $\text{min}[(3d-5)/2,d-1]=d-1$ for all odd $d\geq 3$ instead.

To obtain the above bond dimensions we chose \param{cnotL-cnot-order}=\val{north}. This time the remaining parameters not only affect runtime, but need to be carefully chosen, see Tab.~\ref{tab:msd_bench}---the reasons for this are hard to decipher and probably accidental.
For \param{qubit-order}=\val{north} one may choose, e.g., \param{qec-layer-order}=\val{Z} and \param{qec-cnot-order}=\val{east}.

We note that the logical Hadamard gates include a code rotation which we perform using SWAP gates. Since a SWAP gate requires MPOs with $D=4$ entanglement dimension, we decompose each of them into three CNOTs. While this is not necessarily runtime optimal, it allows us to stick to the $D=2$ chosen for this paper. Note also that SWAPs could in principle be applied virtually through qubit relabeling, leading to another potential optimization~\cite{deger2026efficiently}.

\section{Deep non-Clifford circuits\label{sec:random}}

\begin{figure}
    \centering
    \includegraphics[width=\columnwidth]{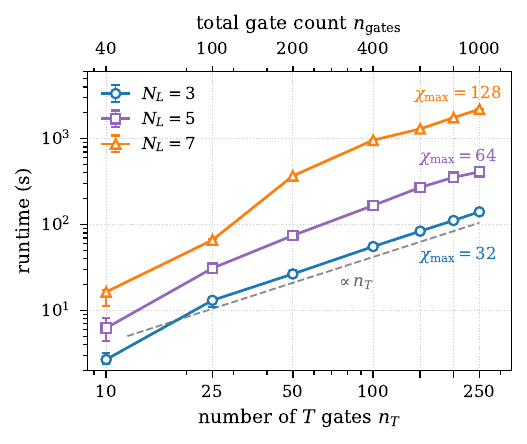}
    \caption{Mean runtimes from random circuits with $N_L\in\{3,5,7\}$ logical qubits encoded in a $d=3$ rotated surface code (). Every qubit starts in either $\ket{0}$ or $\ket{+}$ with 50\% probability. $n_\text{gates}$ random gates drawn from $\{\textsc{H}, \textsc{S}, \textsc{T}, \textsc{CNOT}\}$ are applied with: $n_\text{T}=n_\text{gates}/4$ \textsc{T} gates, and $p_\textsc{CNOT}=0.5$, $p_\textsc{H}=p_\textsc{S}=0.25$ for the remaining $3n_\text{gates}/4$ gates. The bond dimension at the logical level saturates to the maximum of $2^{\lfloor N_L/2\rfloor}$ and at the physical level it is multiplied by 16.}
    \label{fig:random_circuit}
\end{figure}

While the MSD circuit had a moderate number of \textsc{T} gates, increasing the number of \textsc{T} gates in the circuit comes at no extra cost for MPS. This is because single-qubit gates do not create entanglement, which is the relevant metric for MPS.
To demonstrate this we consider as a last example deep random non-Clifford circuits with $N_L$ logical qubits using the rotated surface code of distance $d=3$. While random circuits are known to maximize entanglement and be hard to simulate with MPS, we restrict them to small $N_L$ so the bond dimensions remain under control.

We consider random circuits with each qubit initialized in either $\ket{0_L}$ or $\ket{+_L}$ with 50\% probability. We apply $n_\text{gates}$ random gates drawn from the set \{H, S, T, CNOT\} to arbitrary qubits with: $n_\text{T}=n_\text{gates}/4$ \textsc{T} gates, and $p_\textsc{CNOT}=0.5$, $p_\textsc{H}=p_\textsc{S}=0.25$ for the remaining $3n_\text{gates}/4$ gates. For T and S we apply $\text{T}^\dagger$ and $\text{S}^\dagger$ with 50\% probability instead. For these simulations we used the same physical parameters as for the MSD circuit: \param{qubit-order}=\val{southeast}, \param{qec-layer-order}=\val{Z}, \param{qec-cnot-order}=\val{northeast}, \param{cnotL-cnot-order}=\val{north}; in addition, we use $\lfloor N_L/2 \rfloor$ ancillas symmetrically interspersed. Since the circuit is random and deep we do not use \param{depth-reorder}, nor \param{qubit-reuse}.

Figure~\ref{fig:random_circuit} shows the runtime required to simulate circuits with $N_L\in\{3,5,7\}$ as a function of the number $n_T$ of T gates, or equivalently of total gates. Irrespective of $N_L$, the runtime scales linearly with $n_T$, demonstrating that $T$ gates (and deep circuits) for MPS are cheap (or any non-Clifford one-qubit gate) as long as bond dimensions are constrained. As expected, MPS struggles to simulate large random circuits since $\chi_\text{max}$ increases by a factor of 2 (runtime by 8 because of the $\chi^3$ scaling) with every 2 logical qubits added.

The bond dimension of a random circuit is given by (checked explicitly up to $d=5$)
\begin{equation}
    \chi_\text{max} =  2^{\lfloor N_L/2 \rfloor +d+1}\quad \text{(Random)}.
\label{eq:chimax_random_phys}
\end{equation}
At the logical level the bond dimension saturates at $2^{\lfloor N_L/2\rfloor}$, see also Eq.~(\ref{eq:open_stabs}). The QEC encoding adds a factor of $2^{d+1}$ on top when using \param{qubit-order}=\val{north} (or alternatively, $2^{(3d-1)/2}$ for \param{qubit-order}=\val{southeast}), as explained in the previous section, but without the extra factor of $1/2$ since this is a random state.

\section{Discussion\label{sec:discussion}}

The series of optimizations presented in this paper have allowed us to simulate circuits with hundreds of qubits and \textsc{T} gates with MPS.
For example, given the runtimes of Tab.~\ref{tab:msd_bench} collecting $10^6$ samples for the MSD circuit with $d=3$ would take around $10^4$ computational hours, allowing to estimate moderate logical errors for near-term devices, or acceptance rates.
However, we note that MPS also has the advantage of holding the full state in memory at all times, not just producing samples. This can be useful for directly computing expectation values, and in some cases it can be used to speed up sampling by combining it with efficient Clifford simulators to propagate Pauli noise~\cite{SalesRodriguez2025, sahay2025foldtransversal}.

To put MPS in context, we estimated the difference against other near-Clifford simulation methods for two representative circuits: the MSD $N_L=16$ circuit for $d=3$ and the random circuit with $N_L=5$ and 250 T gates.
We ran these circuits on standard laptop hardware~\footnote{AMD Ryzen 5 7530U, restricted to 4 cores, 22 GiB.} with the  generalized tableau Pauli propagation simulator PPVM~\cite{queraPPVM2026} and our own implementation of a Clifford-Augmented MPS (CAMPS)~\cite{qian2024camps}.
Without in-depth optimization, these two simulators ran the MSD and random circuits in around 7.9~ms / 81~ms (PPVM) and 47~ms / 953~ms (CAMPS), respectively.
Compared to MPS (38.5~s / 410~s), these simulators are around 1000 times faster on these circuits.
We note that we could not run our circuits containing non-Pauli conditional operations with some other recent open-source packages, because they lack the functionality to carry over the state; however, these two results should be representative of similar simulators.

What these methods have in common is that they separate the Clifford and non-Clifford parts of the state.
While the methods to define, represent and compress the non-Clifford part differ, simulation hardness is generally given by the size of this non-Clifford part: for PPVM it's stored as a sparse statevector within a subspace of size $k$ (i.e.~up to $2^k$ basis states; $k$ is also known as `active dimension'~\cite{chaseClifftFastExact2026}), whereas for CAMPS it's an MPS of a given bond dimension.
For our encoded circuits, the active dimension is capped at the number of logical qubits $N_L$ making the circuits simulatable, especially for few \textsc{T} gates as in MSD, but even in the presence of many \textsc{T} gates as in the random circuits with small $N_L$. Similarly, CAMPS can improve on MPS by factoring out the entanglement of the logical qubit states ($\ket{0_L}$, $\ket{1_L}$) into the cheaper Clifford part, leaving only the inter-logical qubit entanglement in the non-Clifford part, whereas bare MPS has to pay the price for both.

While MPS does not show an advantage in these circuits, we show in a follow-up work that MPS is indeed competitive for some recent magic state cultivation circuits~\cite{hartweg2026cultivation}.
Apart from this, there are at least two types of circuits where we expect MPS (and potentially CAMPS) to be beneficial: (1) circuits with low entanglement, high \textsc{T}-gate count, and high active dimension,
for example: a QEC-encoded QFT or a short-time Trotterized Hamiltonian simulation. Both these circuits contain many small-angle rotations and for a few tens of logical qubits $N_L$
we expect that the active dimension will saturate at $N_L$ and MPS will surpass Clifft and PPVM as they hit an exponential wall, whereas MPS remains polynomial in $N_L$ at fixed bond dimension. (2) Circuits with non-Clifford noise, as explored in Ref.~\cite{barone2025colorcode}. In some cases, this noise may not significantly increase entanglement but the active dimension and \textsc{T}-count might increase considerably, as the state exits the logical subspace. Both directions are left for future work.

The circuit optimization methods developed here for reducing the bond dimensions and runtime of MPS can also be used to further improve other near-Clifford methods.
Similar techniques can be directly useful for other tensor network methods such as Tree Tensor Networks~\cite{barone2025colorcode}, as well as recent efforts to combine tensor networks with Clifford simulators~\cite{Masot2024stabilizertensor, mello2024hybridstabsmpos, lami2024quantumstatedesignsclifford, qian2024camps}.
More generally, many of the recently proposed near-Clifford methods~\cite{liSOFTHighperformanceSimulator2025, chaseClifftFastExact2026,tuloupComputingLogicalError2026, queraPPVM2026, fang2026symftuniversal} keep a statevector of coefficients to represent the non-Clifford part of the state, but it constitutes a natural extension to instead use a more efficient MPS representation, which might benefit from similar optimizations adapted to the corresponding Hilbert space.
Finally, the circuit-level optimizations used in this paper lead to orders of magnitude faster runtimes and, even though they were targeted at reducing entanglement, it is possible that related optimizations to reduce, e.g., magic can boost these other methods.
All this might allow to extend such methods beyond their current capabilities, e.g.~in the regime of large active dimensions.

\section{Conclusion}

We have shown that MPS is capable of simulating intermediate-scale QEC circuits including non-Clifford gates with humble computing resources (we used 1 CPU with 4 threads and a few Gb in memory at most)---this is remarkable given that MPS is a general method that is unoptimized to stabilizer(-adjacent) circuits.
Key to the performance achieved was the set of circuit-level optimizations we implemented, which allowed to significantly reduce the necessary bond dimensions and the runtime up to several orders of magnitude compared to naive approaches.
In particular, qubit and gate reordering proved to be key and while some configurations could be explained analytically, a lot of optimizations rely on heuristics and systematic parameter sweeps (see also Ref.~\cite{leonteva2026tuningquantummps}).

While absolute MPS performance on the circuits considered is worse than other recent methods (e.g.~PPVM and CAMPS), performance is problem-dependent and MPS may be useful for other circuits such as cultivation~\cite{hartweg2026cultivation}, large-scale low-entanglement encoded circuits, or non-Clifford noise. 
More generally, this paper offers a `bag of tricks' and enough examples that we hope will help other researchers extend circuit-level optimizations to other circuits and tensor network related methods~\cite{barone2025colorcode, Masot2024stabilizertensor, mello2024hybridstabsmpos, lami2024quantumstatedesignsclifford, qian2024camps}, and possibly to other recent near-Clifford methods with statevector representations which may be substituted by an MPS~\cite{liSOFTHighperformanceSimulator2025, chaseClifftFastExact2026,tuloupComputingLogicalError2026, queraPPVM2026, fang2026symftuniversal}.
Finally, since MPS methods are one of the simplest tensor network methods, are applicable for any gate type, and count with several open-source libraries available, they constitute an attractive complement to these non-Clifford simulation methods for QEC.

\vspace{0.5cm}

\emph{Data availability.}
The circuits underlying Tables~\ref{tab:memory_bench}, \ref{tab:bell_bench} and \ref{tab:msd_bench} and Figs.~\ref{fig:random_circuit}, \ref{fig:memory_runtimes} and \ref{fig:cnot_bonddims} are available at \cite{pineiroorioli2026circuits}. The MPS simulations used MIMIQ\textsuperscript{TM} (QPerfect), which is available under a commercial licence.

\begin{acknowledgments}
We thank Tom Hartweg for collaboration on related work and support with near-Clifford methods, and Johannes Schachenmayer for MPS support. APO would like to thank Laura Pecorari and Hugo Perrin for our motivating QEC therapy sessions, and Marc Serra-Peralta for fruitful discussions around QEC for non-Clifford circuits.
SW acknowledges a state grant managed by the French National Research Agency under the Investments of the Future Program with the reference ANR-21-ESRE-0032 “aQCess - Atomic Quantum Computing as a Service”, the Horizon Europe programme HORIZON-CL4-2021-DIGITAL-EMERGING-01-30 via the project “EuRyQa - European infrastructure for Rydberg Quantum Computing” grant agreement number 101070144 and support from the Institut Universitaire de France (IUF).
\end{acknowledgments}

\bibliography{bib_magic}

\appendix

\section{Memory runtimes}

\begin{figure}[t!]
    \centering
    \includegraphics[width=\columnwidth]{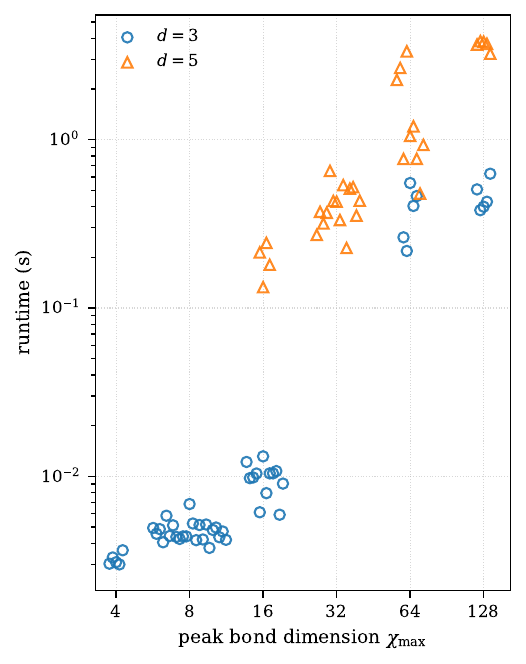}
    \caption{Runtimes and maximal bond dimension $\chi_\text{max}$ of the memory circuit of state $\ket{0_L}$ for $d=3$ and 5 for all the parameters presented in the main text: \param{qubit-order} $\in$ \{\val{naive}, \val{northeast}, \val{southeast}, \val{north}, \val{east}\}, \param{qec-layer-order} $\in$ \{\val{N},\val{Z}\}, \param{qec-cnot-order} same options as \param{qubit-order}. The \val{naive} option was removed from $d=5$. The bond dimensions are all powers of 2 but have been slightly shifted for better visualization. For both distances, the lowest bond dimension results correspond to \param{qubit-order}=\val{northeast} and \param{qec-layer-order}=\val{N}.}
    \label{fig:memory_runtimes}
\end{figure}

We present in Fig.~\ref{fig:memory_runtimes} a scatter plot with all the runtime and $\chi_\text{max}$ results for the memory circuit with distances $d=3$ and 5. Notice that all $\chi_\text{max}$ results are powers of 2, but the points have been slightly shifted for better visualization. For both distances, there are 4 or 5 results that reach the lowest bond dimension of Eq.~(\ref{eq:chimax_memory}) corresponding to the choice \param{qubit-order}=\val{northeast}, \param{qec-layer-order}=\val{N} and any value of \param{qec-cnot-order} tried. The worst runtimes for $d=3$ are comparable to $d=5$ and correspond to the \val{naive} option, which we excluded from the $d=5$ sweep. The plot also shows an approximate $\chi^3$ scaling when moving from $d=3$ to $d=5$. This scaling is less apparent for points within the same distance because of finite-size effects and because cases with larger $\chi_\text{max}$ might only spend few steps at the peak.

\section{Logical CNOT bond dimensions}

\begin{figure}[t!]
    \centering
    \includegraphics[width=\columnwidth]{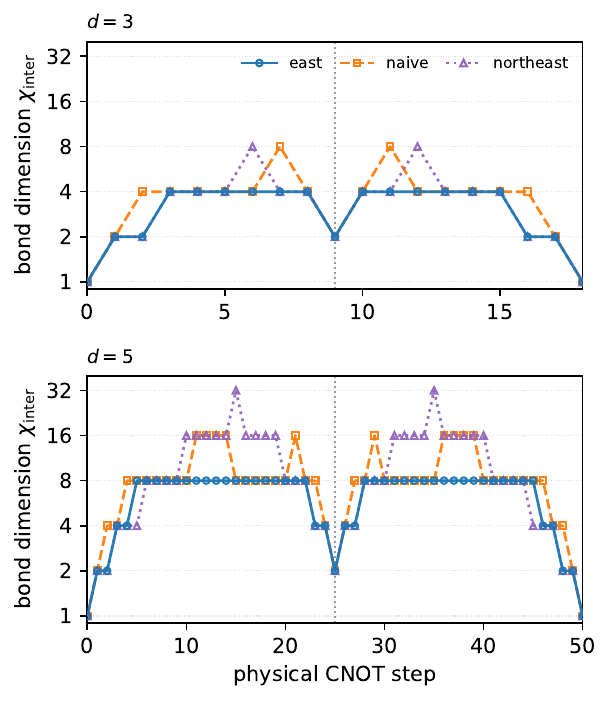}
    \caption{Evolution of the bond dimension $\chi_\text{inter}$ that connects logical qubits 1 and 2 in the Bell state circuit. Each step corresponds to the application of one physical CNOT part of the two logical CNOTs ($d^2$ physical CNOTs each) that are applied to the $\ket{+_L0_L}$ state. We show $d=3$ (top) and $d=5$ (bottom) for three different CNOT orderings: \val{east}, \val{northeast} and \val{naive}. In the middle, a Bell state is created corresponding to $\chi_\text{inter}=2$ and at the end the system returns to a product state $\chi_\text{inter}=1$.}
    \label{fig:cnot_bonddims}
\end{figure}

We present in Fig.~\ref{fig:cnot_bonddims} the evolution of the bond dimension $\chi_\text{inter}$ during application of logical CNOTs in the Bell state circuit. Each step along the $x$-axis corresponds to the application of one physical CNOT belonging to a logical CNOT. The circuit applies two successive logical CNOTs bringing the qubits from $\ket{+_L0_L}$ to a Bell state and back to $\ket{+_L0_L}$, as shown by $\chi_\text{inter}$ going from 1 to 2 to 1. At intermediate steps the bond dimension increases considerably, as discussed in Sec.~\ref{sec:mps_qec}, see the ``mountain'' problem in Fig.~\ref{fig:logstate}. The figure shows that the difference between the peak and the valley grows with distance $d$. The three lines compare three different choices of \param{cnotL-cnot-order}, out of which only \val{east} achieves the optimal bond dimension. Note that the \val{naive} ordering corresponds to the CNOTs ordered in the same way as the qubits of Fig.~\ref{fig:memory}.

\section{Compressed MSD circuits}

\begin{figure}[tb!]
    \centering
    \includegraphics[width=\columnwidth]{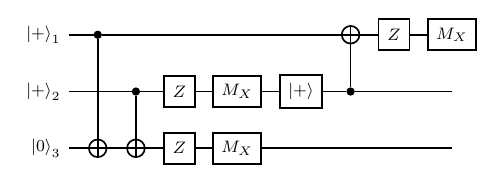}
    \caption{$Z$-gate MSD circuit for $N_L=4$ compressed using \param{depth-reorder}=\val{descending} and \param{qubit-reuse}=\val{true} (see Sec.~\ref{sec:msd}) such that only 3 logical qubits are required. This circuit is equivalent to the one in Fig.~\ref{fig:msd_circuit}a restricted to the first 4 qubits and substituting $T\rightarrow Z$.}
    \label{fig:msd4_compressed}
\end{figure}

Figures \ref{fig:msd4_compressed}, \ref{fig:msd8_compressed} and \ref{fig:msd16_compressed} show the MSD circuit for $N_L=4,8,16$ equivalent to Fig.~\ref{fig:msd_circuit} after it has been optimized to lower the connectivity and bond dimension of the MPS representation using \param{depth-reorder}=\val{descending} and \param{qubit-reuse}=\val{true} (see Sec.~\ref{sec:msd}).

\begin{widetext}

\begin{figure*}
    \centering
    \includegraphics[width=\textwidth]{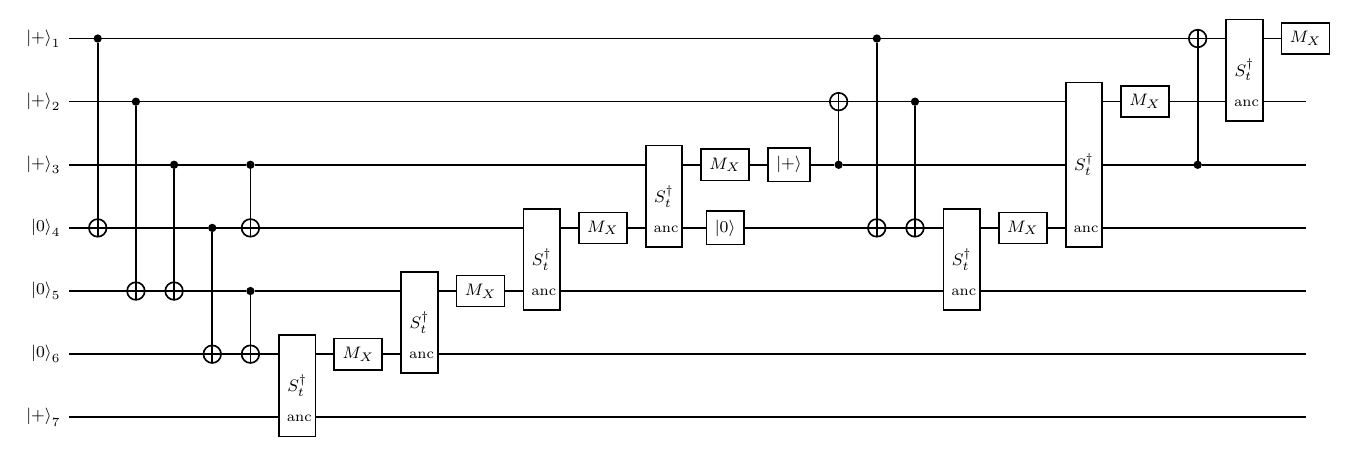}
    \caption{$S$-gate MSD circuit for $N_L=8$ compressed using \param{depth-reorder}=\val{descending} and \param{qubit-reuse}=\val{true} (see Sec.~\ref{sec:msd}) such that only 7 logical qubits are required (including ancilla). This circuit is equivalent to the one in Fig.~\ref{fig:msd_circuit}a restricted to the first 8 qubits and substituting $T\rightarrow S$. The $S^\dagger_t$ boxes correspond to the $S$-gate gadget of Fig.~\ref{fig:msd_circuit}b with the ancilla qubit marked as `anc' and the target qubit placed on the opposite far end of the box.}
    \label{fig:msd8_compressed}
\end{figure*}

\begin{figure*}
    \centering
    \includegraphics[width=\textwidth]{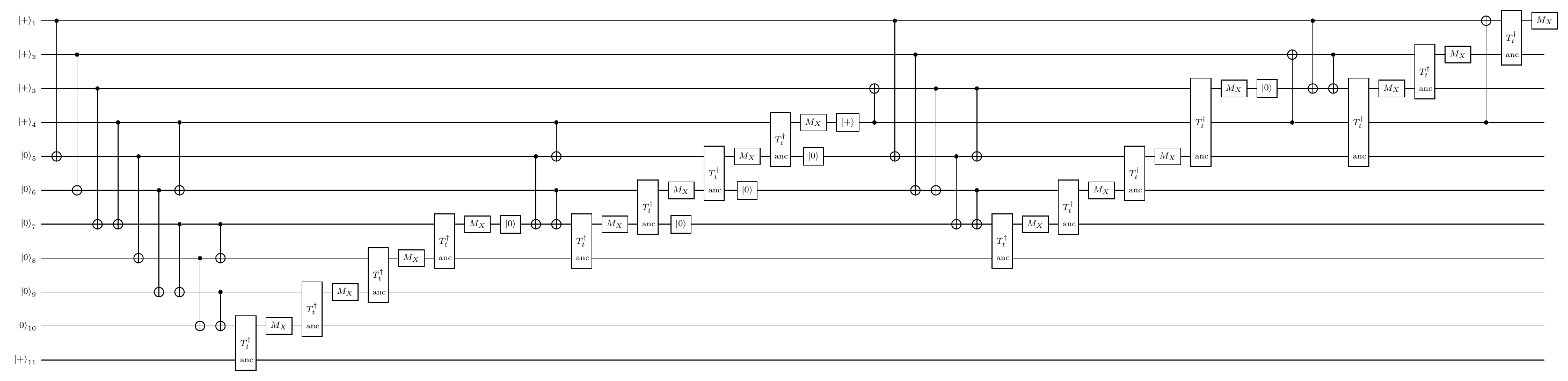}
    \caption{$T$-gate MSD circuit for $N_L=16$ compressed using \param{depth-reorder}=\val{descending} and \param{qubit-reuse}=\val{true} (see Sec.~\ref{sec:msd}) such that only 11 logical qubits are required (including ancilla). This circuit is equivalent to the one in Fig.~\ref{fig:msd_circuit}a. The $T^\dagger_t$ boxes correspond to the $T$-gate teleportation gadget of Fig.~\ref{fig:msd_circuit}b with the ancilla qubit marked as `anc' and the target qubit placed on the opposite far end of the box.}
    \label{fig:msd16_compressed}
\end{figure*}

\end{widetext}

\end{document}